\documentclass{iaus}
\usepackage{natbib}
\usepackage{graphicx}

\title[Opacity in extragalactic sources and astrometry applications] %% give here short title %%
{Opacity in compact extragalactic radio sources and its effect on
radio-optical reference frame alignment}

\author[Y.~Y.~Kovalev et al.]   %% give here short author list %%
{
Y.~Y.~Kovalev$^{1,2}$,
A.~P.~Lobanov$^1$,
A.~B.~Pushkarev$^{1,3,4}$,
\and
J.~A.~Zensus$^1$
}

\affiliation{
$^1$Max-Planck-Institut f\"ur Radioastronomie,
Auf dem H\"ugel 69, 53121 Bonn, Germany \\
e-mail: {\tt 
ykovalev, alobanov, apushkar, azensus@mpifr-bonn.mpg.de}
\\[\affilskip]
$^2$Astro Space Center of Lebedev Physical Institute,\\
Profsoyuznaya 84/32, 117997 Moscow, Russia\\[\affilskip]
$^3$Pulkovo Astronomical Observatory, Russia; $^4$Crimean Astrophysical Observatory, Ukraine
}

\pubyear{2008}
\volume{248}  %% insert here IAU Symposium No.
\pagerange{1--4}
\jname{A Giant Step: from Milli- to Micro-arcsecond Astrometry}
\editors{W.-J.~Jin, I.~Platais, M.~Perryman, eds.}
\begin{document}

\maketitle

\begin{abstract}
Accurate alignment of the radio and optical celestial reference frames
requires detailed understanding of physical factors that may cause
offsets between the positions of the same object measured in different
spectral bands. Opacity in compact extragalactic jets (due to
synchrotron self-absorption and external free-free absorption) is one of
the key physical phenomena producing such an offset, and this effect is
well-known in radio astronomy (``core shift''). We have measured the
core shifts in a sample of 29 bright compact extragalactic radio sources
observed using very long baseline interferometry (VLBI) at 2.3 and 8.6
GHz. We report the results of these measurements and estimate that the
average shift between radio and optical positions of distant quasars
would be of the order of 0.1--0.2 mas. This shift exceeds positional
accuracy of GAIA and SIM.
We suggest two possible approaches to carefully investigate and
correct for this effect in order to align accurately the radio and
optical positions.  Both approaches involve determining a Primary
Reference Sample of objects to be used for tying the radio and optical
reference frames together.

\end{abstract}

\section{Introduction \label{s:intro}}

Extragalactic relativistic jets are formed in the immediate vicinity of
the central black holes in galaxies, at distances of the order of 100
gravitational radii, and they become visible in the radio at distances
of about 1000 gravitational radii \citep{LZ2007}. This apparent origin
of the radio jets is commonly called the ``core''. In radio images of
extragalactic jets, the core is located in the region with an optical
depth $\tau_s\approx 1$. This causes the absolute position of the core,
$r_\mathrm{core}$, to vary with the observing frequency, $\nu$, since the
optical depth profile along the jet depends on $\nu$: $r_\mathrm{core}
\propto \nu^{-1/k_\mathrm{r}}$ \citep{BlandfordKonigl79}.
Variations of the
optical depth along the jet can result from synchrotron self-absorption
\citep{Koenigl81}, pressure and density gradients in the jet and
free-free absorption in the ambient medium most likely associated with
the broad-line region (BLR) \citep{L98}.

The core shift is expected to introduce systematic offsets between the
radio and optical positions of reference sources, affecting strongly the
accuracy of the radio-optical matching of the astrometric catalogues.
The magnitude of the core shift can exceed the inflated errors of the
radio and optical positional measurements by a large factor. This makes
it necessary to perform systematic studies of the core shift in the
astrometric samples in order to understand and remove the contribution
of the core shift to the errors of the radio-optical position
alignment.

Measurements of the core shift have been done so far only in a small
number of objects
\citep[e.g.,][]{MES94,Lara_etal94,PR97,L96,L98,RL2001,Kadler_etal04,SokolovCawthorne2007}.
In this paper, we present results for 29 compact extragalactic radio
sources used in VLBI astrometric studies and discuss the core shift
effect on the task of the radio-optical reference frame alignment.

\section{Core shift measurements between 2.3 and 8.6~GHz\label{s:res}}

%%%%%%%%%%%%%%%%%%%%%%%%%%%%%%%% Figure
\begin{figure}[b]
\begin{center}
\resizebox{0.6\hsize}{!}{
   \includegraphics[trim=0cm 0cm 0cm 0cm]{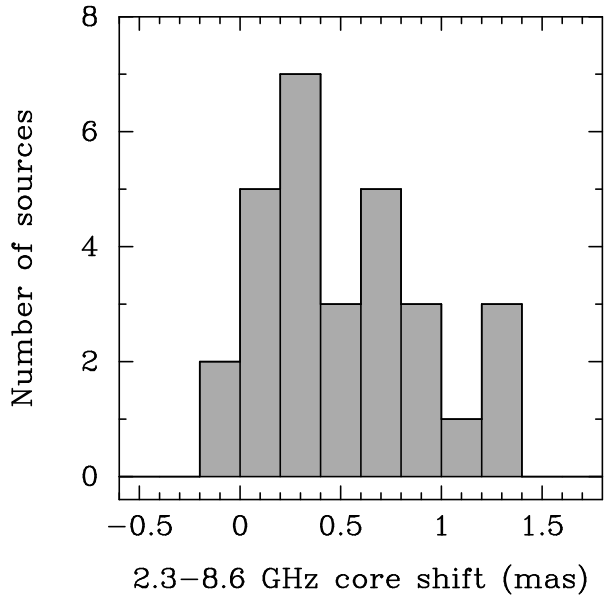}
}
\end{center}
\caption{
\label{f:hist_shifts}
Histogram of the derived core shift values for 29 sources.
One average core shift value per source is used.
The median value for the distribution is equal to 0.44~mas.
}
\end{figure}
%________________________________________________________________

We have imaged and analyzed 277 sources from ten 24\,hr-long geodetic
RDV (Research and Development VLBI experiments, see, e.g.,
\citealt*{Gordon05}) observations made in 2002 and 2003. Geodetic RDV
sessions feature simultaneous observations at 2.3~GHz and 8.6~GHz (S and
X bands) with a global VLBI network which includes, for every session, the
VLBA and up to nine other radio telescopes around the world.

This long-term RDV program is one of the best choices for a
large project to measure two-frequency core shifts for several
reasons: (i) it is optimized to have a good ($u$,$v$)-coverage, (ii) it
has the maximum possible resolution for ground-based VLBI at these
frequencies, (iii) the frequency ratio between the simultaneously
observed bands is high (3.7), and (iv) the core shift per unit of
frequency between 2.3 and 8.6~GHz is larger than that at higher
frequencies \citep[see, e.g.,][]{L98}.
A dedicated multi-band VLBI project covering a wider frequency range
would certainly provide results of a better quality, but is extremely
time consuming.

We have measured the frequency-dependent core shift between 2.3 and 8.6
GHz by model-fitting the source structure with two-dimensional Gaussian
components \citep{pearson1999} and referencing the position of the core
component to one or more jet features, assuming the latter to be
optically thin and having frequency-independent peak positions.
The shifts were successfully measured 
in 29 Active Galactic Neclei (AGN),
with the determined values of the shift ranging between $-0.1$ and
1.4~mas and the median value for the sample of 0.44~mas
(Fig.~\ref{f:hist_shifts}). Typical errors of the core shift
measurements in this study are about or less than 0.1~mas. For 90\% of the 277
objects imaged, no reliable estimates of core shifts have been obtained
by this method.

\section{Radio-optical alignment of astrometric positions\label{s:radio-optics}}

The core shift issue gains specific importance when the
radio reference frame needs to be connected to an optical reference
frame. So far the link is based on the study of some radio stars which
are seen both by {\em Hipparcos} and VLBI. These measurements
are based on stars with large radiospheres and with big spots at the
optical surfaces; both effects may introduce large uncertainties
\citep[see, e.g.,][]{Lestrade_etal95,Boboltz2005,Ros2005}. Future
accurate alignment of the frames has to rely on using compact radio
sources in distant quasars.

Below, we discuss the alignment problem for compact extragalactic
radio sources. We assume that the dominating component in both the radio
and optical bands is the synchrotron self-absorbed compact jet origin
(core). Broad-band modeling of blazar spectral energy distribution
supports this hypothesis \citep[see, e.g., recent review by][]{B07}.
High-resolution VLBI observations of nearby AGN imply that the jet is
formed and emitting in the radio already at distances of $\le 1000$
gravitational radii from the central engine
\citep[e.g.][]{Junor_etal99,Kadler_etal04}. Thus the physical offset
between the jet base and the central nucleus can be much smaller than the
positional shift due to opacity in the jet (the latter can be larger
than 1 pc). This implies that the offset between radio and optical
positions of reference quasars will be dominated by the core shift even
if the optical emission comes from the accretion disk around the central
nucleus.

The magnitude of the core shift, $\Delta r$, between two arbitrary frequencies
$\nu_1$ and $\nu_2$ ($\nu_1 > \nu_2$) caused by synchrotron
self-absorption can be predicted for an object with known synchrotron
luminosity, $L_\mathrm{syn}$, of the compact jet \citep{L98}.
If not corrected for, the core shift will introduce an additional
additive error factor in the alignment of the radio and optical
reference frames. 
For typical parameters of relativistic jets we estimate an average shift
between 8.6\,GHz and 6000\,\AA\ for a complete sample of compact extragalactic
sources to be of an order of 0.1--0.2~mas.

\section{Summary \label{s:sum}}

Measurements of the frequency-dependent shift of the parsec-scale jet
cores in AGN are reported for 29 bright extragalactic radio sources. It
is shown that the shift can be as high as 1.4~mas between 2.3 and
8.6~GHz. We have shown that core shifts are
likely to pose problems for connecting radio and optical reference frames. 
We have estimated from theory an average shift between the radio (4~cm)
and optical (6000~\AA) bands to be of an order of 0.1~mas for a
complete sample of radio selected AGN.

The estimated radio-optical core shift exceeds the positional accuracy
of GAIA and SIM. It implies that the core shift effect should be
carefully investigated, and corrected for, in order to align accurately
the radio and optical positions. Based on our investigation, we suggest
two possible approaches, both involving determining a Primary Reference
Sample of objects to be used for tying the radio and optical reference
frames together.
1)
In the first approach, multi-frequency VLBI measurements can be used for
calculating the projected optical positions, assuming that the radio and
optical emission regions are both dominated by a spatially compact
component marginally resolved with VLBI and SIM and point-like for GAIA.
The discrepancies between the measured optical and radio positions can
then be corrected for the predicted shifts, and the subsequent alignment
of the radio and optical reference frames can be done using standard
procedures. 
2)
A more conservative approach may also
be applied, by employing the VLBI observations to identify and
including in the Primary Reference Sample only those quasars in which
no significant core shift has been detected in multi-epoch
experiments. Either of the two approaches should lead to substantial
improvements of the accuracy of the radio-optical position alignment.

\begin{acknowledgements}
These proceedings are based on a paper by Y.~Y.~Kovalev et al.~(A\&A,
2007, submitted).
The raw VLBI data were provided to us by the open NRAO archive. The
National Radio Astronomy Observatory is a facility of the National
Science Foundation operated under cooperative agreement by Associated
Universities, Inc.
     Y.~Y.~Kovalev is a Research Fellow of the Alexander von Humboldt
Foundation. Y.~Y.~Kovalev was supported in part by the Russian
Foundation for Basic Research (project 05-02-17377) while working in
Moscow in the first half of 2006.
     We would like to thank Patrick Charlot,
Ed Fomalont, Leonid Petrov, Richard Porcas,
Eduardo Ros as well as the NASA GSFC VLBI group and the MOJAVE team for
fruitful discussions.
%     This research has made use of the NASA/IPAC Extragalactic Database (NED)
%which is operated by the Jet Propulsion Laboratory, California Institute
%of Technology, under contract with the National Aeronautics and Space
%Administration.
%     This research has made use of NASA's Astrophysics Data System.
\end{acknowledgements}

%\bibliographystyle{apj}
%\bibliography{yyk}

\end{document}